\documentclass[onefignum, onetabnum,onealgnum,onethmnum]{siamonline190516}
\usepackage{comment}
\usepackage{lipsum}
\usepackage[margin=1in, footskip=36pt]{geometry}

\usepackage{endnotes}

\usepackage{titlesec}
\usepackage{titletoc}

\titleclass{\task}{straight}[\section]
\newcounter{task}
\renewcommand{\thetask}{\arabic{task}}
\titleformat{\task}[hang]
    {\normalfont\LARGE\bfseries}{Task \thetask:}{1em}{}

\titleformat*{\task}{\color{header1}\bfseries}

\titlecontents{task}
              [3.8em] 
              {}
              {\contentslabel{2.3em}}
              {\hspace*{-2.3em}}
              {\titlerule*[1pc]{.}\contentspage}

\titlespacing*{\section}{0ex}{1ex}{1ex}
\titlespacing*{\subsection}{0ex}{1ex}{1ex}
\titlespacing*{\paragraph}{0ex}{1ex}{1ex}
\titlespacing*{\subparagraph}{0pt}{1ex}{1ex}
\titlespacing*{\task}{0em}{1ex}{1ex}

\usepackage[sort&compress,comma,square,numbers]{natbib}

\usepackage{amsfonts,amsmath,amssymb}
\usepackage{bbm}

\usepackage{multicol}
\usepackage{enumitem}
\setlist[enumerate]{noitemsep}
\setlist[itemize]{noitemsep}
\setlist[description]{noitemsep}

\usepackage{hhline}

\usepackage{todonotes}
\RequirePackage{ulem}

\usepackage{booktabs}
\usepackage{titlesec}
\usepackage{fancyhdr}
\usepackage{fullpage}

\RequirePackage{titlesec}
\usepackage[T1]{fontenc}
\usepackage[scaled]{helvet}
\usepackage[scaled=1.10, med]{zlmtt}

\usepackage{algorithm}
\usepackage{algpseudocode}

\newcommand{\argmax}{\operatornamewithlimits{argmax}}

\newcommand{\del}{\delta}

\usepackage{math}

\title{
\Large
Statistical Analysis of Data Repeatability Measures}

\author{Zeyi Wang$^1$, Eric Bridgeford$^1$, Shangsi Wang$^1$, Joshua T. Vogelstein$^{1,2}$, Brian Caffo$^1$}

\usepackage{xcolor}
\usepackage{soul}

\begin{document}

\maketitle

\begin{abstract}
The advent of modern data collection and processing techniques has seen the size, scale, and complexity of data grow exponentially. 
A seminal step in leveraging these rich datasets for downstream inference is understanding the characteristics of the data which are repeatable 
--- the aspects of the data that are able to be identified 
under duplicated analyses.
Conflictingly, the utility of traditional repeatability measures
 , such as the intraclass correlation coefficient, 
under these settings is limited.
In recent work, novel data repeatability measures have been introduced 
in the context where a set of subjects are measured twice or more, including: fingerprinting, rank sums, and generalizations of the intraclass correlation coefficient.  
However, the relationships between, and the best practices among these measures remains largely unknown. 
In this manuscript, we formalize a novel repeatability measure, discriminability. 
We show that it is deterministically linked with the intraclass correlation coefficients under univariate random effect models, and has the desired property of optimal accuracy for inferential tasks using multivariate measurements.
Additionally, we overview and 
systematically compare existing repeatability statistics with discriminability, using both theoretical results and simulations. 
We show that the rank sum statistic is deterministically linked to a consistent estimator of discriminability.  
The statistical power of permutation tests derived from these measures are compared numerically under Gaussian and non-Gaussian settings, with and without simulated batch effects. Motivated by both theoretical and empirical results, we provide 
methodological recommendations for each benchmark setting to serve as a resource for future analyses. We believe these recommendations will play an important role 
towards improving repeatability in fields such as functional magnetic resonance imaging, genomics, pharmacology, and more.    
\end{abstract}


\let\thefootnote\relax\footnote{


\begin{@abssec}{Author Information and Acknowledgements}{$^1$ Johns Hopkins University, $^2$ Progressive Learning. } 
\end{@abssec}
}



\section{Introduction}

\textit{Data repeatability} is defined as consistency or similarity across technical replicates of a measurement.  We emphasize that we use the term without knowledge of, or assuming that one of the replicates is the correct, true measurement. The same definition is often referred to as \textit{test-retest reliability}, or the reproducibility of a measurement procedure, where the consistency of repeated measurements is being emphasized \citep{muller1994critical}. However, note that the general concepts of reliability and reproducibility are often applied beyond the definition of consistency in repeated measurements, depending on the context. General reviews of the concept of research reproducibility, with comparison to replicability can be found in \cite{goodman2016does, patil2016statistical}. A rich literature exists for other related, but distinct, types of reliability, such as inter-rater reliability (an overview can be found \cite{gwet2014handbook}). In summary, we selectively focus on the evaluation of data repeatability, as a crucial starting point for evaluating measurement validity.

Data repeatability reflects the stability of the entire data generating process, which often introduces inevitable noise and variability \citep{bach2012knowing, beck2012not}, or potentially involves complex data collection and processing, especially for modern large data studies \citep{labrinidis2012challenges, garcia2016big, lichtman2014big}. Data repeatability can be considered the counterpart of what is referred to as stability in the statistical methodology literature \citep{yu2013stability}, an important component of scientific reproducibility. 
The critical role for data repeatability has become highlighted in the light of the so-called ``reproducibility crisis'' discussed across many scientific domains \citep{Baker2016-BAKSL-2, open2015estimating, button2013power}. 
Repeatability is also used as a key tool to detect likely irreproducible findings and statistical errors. For example, a recent criticism over issues in the repeated use of data in the field of cognitive neuroscience \citep{vul2009puzzlingly} relied on the absence of repeatability as proof of the issue. Some have also argued that the misinterpretation of repeatability can result in false confidence in a study's reproducibility and subsequently leading to the neglect of important design issues \citep{turner2018small}. We conclude that a thorough investigation and accurate interpretation of data repeatability is crucial for a better understanding of existing issues of reproducibility and working towards better future practices. 

The intraclass correlation coefficient (ICC) is a commonly used metric for data repeatability or test-retest reliability. However, the ICC is limited in several ways when applied to multivariate and complex big data. 
The definition and inference of ICC is often based on a relatively strict parametric ANOVA model that assumes separability and additivity. Typically, Gaussian assumptions are applied for inference. 
Various ICC generalizations have been proposed to partially relax Gaussian assumptions or to incorporate rank-based definitions, thus improving robustness against model misspecification. However, ICC and its generalizations apply to univariate data, and there is no consensus on how one should synthesize multiple (generalized) ICC's over dimensions of multivariate measurements, or for measurements with varying dimensions. 

Recently, several novel data repeatability measures have been proposed, including fingerprinting \citep{finn2015functional, finn2017can, wang2018statistical}, rank sums \citep{airan2016factors}, and the image intraclass correlation coefficient (I2C2) \citep{shou2013quantifying}, which is a generalization of the classical univariate ICC. Unlike univariate methods, such as ICC, these newly proposed methods can handle high-dimensional complexity and computationally scale. Moreover, by building the measures on rank transformations, the nonparametric methods (fingerprinting, rank sums) are robust to variations in parametric assumptions.  However, the relationships between these methods and the best practices among them remain largely unknown. Furthermore, relationships in interpretation and performance have not been established. Thus, often less effective or robust measures of data quality are being used, potentially leading to worse study practices, worse processing pipelines and sub-optimal performance of prediction algorithms.  

In this manuscript, we particularly focus on discriminability \citep{bridgeford2021eliminating}, a new data repeatability measure. Discriminability is defined on a general repeated measurement model that is free of parametric assumptions, yet remains deterministically linked to ICC for univariate measurements when ANOVA assumptions are met.  It has been proven that the most discriminable measurements are optimally accurate in the Bayes error rate of subsequent inferential tasks, regardless of what the actual task is.  While discriminability is a promising multivariate and non-parametric measure of data repeatability, its statistical properties and relationships to other measures remains largely unstudied.

We focused on discriminability by investigating its mathematical relationships with other multivariate repeatability measures, regardless of parametric restrictions. This resembles the relation between optimal intra- and inter-subject correlations and ICC of the measurements under univariate scenarios, an idea that has been recently studied and discussed in neuroimaging \citep{vul2009puzzlingly, bennett2010reliable, zuo2014test}. 

In addition, we numerically compared the methods in the terms of their ability to detect significance in permutation tests specifically designed to detect the presence of data repeatability. To summarize, our results illustrate the general power advantages of discriminability, when compared to other nonparametric methods, and its robustness advantages against violation of Gaussian assumptions, when compared to parametric methods. Of course, parametric methods may be more powerful when distributional assumptions are satisfied. In addition, the rank sum method shows additional robustness against mean shift batch effects compared to discriminability. 

The field of functional magnetic resonance imaging (fMRI) is a concrete example where data repeatability is of key interest, because the data quality can be impacted by noise, biological confounds, and complex acquisition and processing choices. 
The results presented in this manuscript have the potential to improve the evaluation and optimization of fMRI data repeatability. General reviews of fMRI reliability can be found \cite{bennett2010reliable, frohner2017addressing}, although emphasis was put on the reliability of results, not necessarily restricted to measurement.
For example, popular cluster-overlap-based reliability measures, such as the Dice coefficient and Jaccard index, are restricted to the similarity analysis of graphs and sets and are not applicable for other types of data. 
Some concepts similar to inter-rate reliability in fMRI, such as inter-site, inter-scanner or inter-technologist reliability, are not discussed in detail, but the measures discussed in this manuscript can be applied. 

The quantification of data repeatability is even more crucial for fMRI-based functional connectivity (FC), where second order statistics (usually correlations) or various network-based graph metrics are the object of study. Resting state correlations are particularly sensitive to biological confounds, in contrast to task based fMRI, where the confound is often not correlated with the task. Variability can be induced by changes in physiological and cognitive status of a subject, within a single scan session or between two sessions that are hour, days, or months apart. 
In addition, common practices in the field can raise questions of data quality \citep{zuo2014test, jiang2015toward}. For instance, autocorrelations in the BOLD time series might violate independence and parametric assumptions in correlation analyses. Averaging the time series over a large region may involve voxels with low functional homogeneity and introduce spurious variability. It is also a concern when, as is typical, a number of reasonable preprocessing options are available that produce varying measurement outcomes. Processing choices can be particularly difficult to generalize across studies, since target measurements can be on different scales or formed with a different data reduction strategy (seed-to-voxels, voxel-by-voxel, region-by-region, etc.). In all these scenarios, understanding data repeatability is prerequisite for any meaningful scientific discovery or clinical application. Objective repeatability measures, preferably non-parametric and able to accommodate varying data dimensions, (such as the nonparametric measures discussed in this manuscript) are needed. Moreover, data repeatability has uses beyond assessing the quality of correlation-based FC measurements \citep{noble2019decade}, but is also used in many other settings. Some examples include: $i.$ selecting best practices for data acquisition and preprocessing \citep{pervaiz2019optimising}, $ii.$ identifying FC biomarkers \citep{gabrieli2015prediction, castellanos2013clinical, kelly2012characterizing}, $iii.$ optimizing FC-based prediction models \citep{svaldi2019optimizing}, and $iv.$ evaluating the accuracy of multi-class prediction algorithms \citep{zheng2018extrapolating}. 

\section{Review of Existing Data Repeatability Measures} \label{sec:repeatability}

In this section, we will define several measures of data repeatability under associated statistical models. We emphasize that most repeatability measures are not directly associated with models or likelihoods. For example, the intraclass correlation coefficient can be motivated via moments or a Gaussian random effect model. Alternatively, fingerprinting and rank sums are purposefully designed without parametric assumptions. Thus, some repeatability measures do not have population analogs, i.e. they're estimators without clear estimands. In what follows, we hope to clarify the population based estimands as well as performance under known modeling assumptions.

\subsection{Intraclass Correlation Coefficient}

We consider two types of intraclass correlations, ICC and I2C2 \citep{shou2013quantifying}. Without modification, ICC is designed to evaluate the repeatability of one dimensional measurements. Non-unique generalizations can be made for multivariate measurements, for example, by averaging ICCs over each of the dimensions or by counting percentage of dimensions that pass a threshold on ICC. However, for the latter scenario there is no consensus on best practices, and the interpretation may be subjective to the researcher's choices.
I2C2 generalizes to higher dimensions by providing a multivariate model that decomposes variation into a sum of intra- and inter-subject levels and defines the I2C2 of the fraction of total variation that is total inter-subject variation.  Other generalizations of ICC not discussed further include two-way ANOVA models \citep{shrout1979intraclass} and variations on the Alpha and Kappa statistics. These are not relevant for the evaluation of test-retest reliability \citep{rousson2002assessing, bruton2000reliability}. 

Suppose that we have $n$ subjects, each with $s$ measurements. 
For $l$-dimensional measurements,  
throughout we refer to the following Gaussian random effect model as a Multivariate Analysis of Variance (MANOVA) model:
\begin{equation}
\pmb{x}_{it} = \pmb\mu + \pmb\mu_i + \pmb e_{it}, 
\label{eq:manova}    
\end{equation}
where $\pmb\mu_i \distas{iid} \Norm[l]{\pmb 0, \pmb{\Sigma}_{\pmb  \mu}}$, $\pmb e_{it} \distas{iid} \Norm[l]{\pmb 0, \pmb{\Sigma}}$, independently. (All of the vectors are $l$-dimensional.) Note that $l=1$ leads to the univariate Analysis of Variance (ANOVA) model as a special case. 

In the univariate case of (\ref{eq:manova}), ICC is defined as:
\begin{equation*}
    \lambda = \corr{x_{it}, x_{it'}} = \frac{\sigma_\mu^2}{\sigma_\mu^2 + \sigma^2},
\end{equation*}
for all $t' \not= t$. Assuming the measurements of a same subject form a class, then $x_{it}$ and $x_{it'}$ are both from the $i$-th class, hence the name ("intra-class").  For the multivariate case (\ref{eq:manova}), a  generalization of ICC uses:
$$
\frac{h(\pmb \Sigma_{ \mu})}{h(\pmb \Sigma_{\mu}) + h(\pmb \Sigma)}
$$
where $h$ is a univariate summary of variance covariance matrices. Specifically, Wilk's lambda, which we will refer to as $\Lambda$, uses the determinant, whereas other generalizations, such as I2C2 use the trace, henceforth referred to as $\Lambda_{tr}$.

This repeatability measure is particularly useful for high-dimensional imaging settings and was utilized in the the image intraclass correlation coefficient (I2C2) \citep{shou2013quantifying}. Recall that the trace of the covariance matrix captures the total variability of the random quantity of interest. Then, $\Lambda_{tr}$ intuitively represents the fraction of the variability in the observed data $\pmb x_{it}$ due to the subject effect, $\pmb \mu_i$.

For the univariate case, the estimation of ICC is often conducted through a one-way ANOVA. It is also well known that $\hat\lambda = (F - 1)(F - 1 + s)$ is a non-decreasing function of the F statistic given $s \geq 2$.  I2C2 builds on
this strategy using a hierarchical generalization on principal components called multilevel functional principal components analysis (MFPCA) \citep{di2009multilevel}. 
The MFPCA algorithm utilizes a moment based approach to separate variability into inter- and intra-subject components in a method similar to Henderson's equations in mixed models \citep{henderson1959estimation}. 
Singular value decomposition tricks can be used to make calculations tractable in high dimensions \citep{zipunnikov2011multilevel}. 
In principle, other multivariate approaches can be used to estimate $\Lambda_{tr}$ and $\Lambda$. In addition, latent Gaussian models \citep{chib1998analysis} can extend these approaches to binary data and graphs \citep{yue2015estimating}. 

One of the commonly discussed properties of ICC is its relation with the optimal correlation between two univariate outcomes \citep{vul2009puzzlingly, bennett2010reliable, zuo2014test}. It states: 
\begin{align*}
    \corr{x^{(1)}_{it}, x^{(2)}_{it}} = \corr{\mu^{(1)}_i, \mu^{(2)}_i} \sqrt{\textrm{ICC}(x^{(1)}_{it}) \cdot \textrm{ICC}(x^{(2)}_{it})}, 
\end{align*}
where $x^{(1)}_{it}$ and $x^{(2)}_{it}$ follow the ANOVA model with subject random effects as $\mu^{(1)}_{i}$ and $\mu^{(2)}_{i}$ respectively. This relationship holds without Gaussian distributional assumptions. 
Thus, all else equal, $x^{(1)}_{it}$ and $x^{(2)}_{it}$ are more correlated the higher the individual ICCs.

Previous works have investigated generalizations of univariate ICC that relax the normality assumption on subject effects $\pmb \mu_i$ \citep{ukoumunne2002comparison} or use rank-based definition \citep{hunsberger2022rank} with underlying copula \citep{romdhani2014exchangeable,genest2011estimators}. The rank-based measures are in general robust against marginal misspecifications. However, it remains as a challenge the joint evaluation of multiple univariate ICC's (or generalized ICC's) over a large number of dimensions of complex multivariate data or data with varying dimensions. In this manuscript, for the purpose of evaluating data repeatability of potentially high-dimensional data generating process, we will focus on ICC and I2C2 and leave other generalizations of ICC for future research.

\subsection{Fingerprinting}
\label{sec:fpt}

Fingerprinting is the idea of to matching individuals within a group of subjects at one time to the same group at another time and comparing the frequency by which subjects match to themselves versus others \citep{wang2018statistical}.  The count or proportion of correct matches for a matching scheme represents an intuitive summary of data repeatability. This measure has become especially popular in neuroimaging due to its intuitive appeal \citep{anderson2011reproducibility, finn2015functional, xu2016assessing}.

We first formalize the idea of a population-level fingerprinting measure for repeated measurements. It is assumed that each subject is measured twice, and that the measurement is possibly multivariate.  Then each subject, $i$, at time point, $t$, has measurement, $\pmb{x}_{it}$, $i = 1, \dots, n$, $t = 1, 2$. Assume a distance metric, $\delta(\cdot, \cdot)$, defined between measurements and let 
$\delta_{i,1,2} = \delta(\pmb{x}_{i1}, \pmb{x}_{i2})$, and $\delta_{i, i', 1, 2} = \delta(\pmb{x}_{i1}, \pmb{x}_{i'2})$. 
Define the population level fingerprint index as: 
\begin{align}
    F_{index} = \prob{\delta_{i,1,2} < \delta_{i,i',1,2}; \;\forall i' \neq i}, \label{eq:FI}
\end{align}
where the probability is calculated over a random sample of $n$ subjects. This is the population probability that a random subject matches themselves over any other in the sample. As far as we know, this is the first instance defining a population analog to the sample fingerprinting statistic.

This population measure is implicitly defined without any parametric assumptions. For (\ref{eq:FI}) to be a meaningful population quantity, it is required that the $F_{index}$ is equal for all $i$ and $i$'s, which covers the (M)ANOVA models 
(\ref{eq:manova}) with Gaussian random effects as special cases. More generally, 
random sampling and exchangeability assumptions will also satisfy this requirement. No direct relationship
between the population ICC and this measure, similar to the lack of a direct relationship between the associated sample quantities.

The natural estimate of \eqref{eq:FI} is the proportion of correct matches in a group of subjects. This requires assuming a matching strategy, such as whether matching is done with or without replacement \citep{wang2018statistical}. Almost all fingerprint index studies use matching with replacements as follows. The total number of correct matches (with replacement) is $T_n = \sum_{i = 1}^n \indicator{\delta_{i,1,2} < \delta_{i,i',1,2}; \;\forall i' \neq i}$, where $\indicator{\cdot}$ is the indicator function. Then, the fingerprint index estimator is simply the proportion of correct matches, $\hat{F}_{index} =\frac{T_n}{n}.$

\subsection{Rank Sums} In the test-retest setting with $s = 2$, the fingerprint statistic can be generalized as a Mann-Whitney style statistic.  Instead of counting the events where $\pmb{x}_{i2}$ is the closest to $\pmb{x}_{i1}$ when compared to all other $\pmb{x}_{i'2}$ (for $i' \not = i$), consider calculating the rank. Then the fingerprinting index would simply be the proportion of instances when the rank equals 1.  Formally, let $r_{i1}, r_{i2}, ...,r_{in}$ be the rank transformation of  $\delta_{i,1,1,2}, 
\delta_{i,2,1,2}, ... \delta_{i,n,2,1}$. Then, the rank sum statistic is defined by simply summing up the $r_{ii}$; that is, the ranks of $\delta_{i, 1, 2}$ among all $\delta_{i, i', 1, 2}$. Assuming that there are no ties, the rank sum statistic is defined as:
\begin{align}
R_n = \sum_{i = 1}^n r_{ii} = \sum_{i=1}^n \sum_{i' \not = i} \indicator{\delta_{i, 1, 2} < \delta_{i, i', 1, 2}}. \label{eq:rs}
\end{align}
As mentioned,  the ranks are sufficient for determining the fingerprint index; notationally, $\indicator{\delta_{i,1,2} < \delta_{i,i',1,2}; \;\forall i' \neq i} = \indicator{r_{ii} = 1}$. Of course, the fingerprinting statistic ignores the information contained in ranks, other than the number of the ranks equal to $1$ within subjects. Thus, it may seem obvious that the rank sum statistic is superior to the fingerprint statistic in some sense. However, it should also be noted that the rank sum statistic lacks an intuitive relationship with a population quantity, like the fingerprint statistic does with the fingerprint index.  In addition, both the fingerprint and rank sum statistics lack an obvious generalization for repeated measurements, as they were developed on paired measurements.

\section{Discriminability as a Repeatability Measure}
In this section, we formally define the concept of discriminability under multivariate repeated measurements. We will prove that discriminability is a repeatability measure by being deterministically related to ICC when the classical parametric assumptions are met. Notably, an optimal accuracy property of discriminability in the Bayes error rate is applicable for multivariate measurements without requiring strong parametric assumptions. 
We will also investigate the relation between discriminability and the other measures with the goal of increasing interpretability across studies when using different repeatability measures. 

\subsection{General Model of Repeated Measurements}
\label{sec:setting}
Let $\pmb v_i \in \pmb {\mathcal V}$ be a true physical property of interest for subject $i$. Without the ability to directly observe $\pmb v_i$, we instead observe $\pmb w_{it} = f_\phi(\pmb v_i, t)$, for some random measurement process $f_\phi: \pmb {\mathcal V} \times T \rightarrow \pmb{\mathcal W}$, where $\phi \in \pmb \Phi$ characterizes the measurement process, and $\pmb w_{it} \in \pmb{\mathcal W}$ is the observed measurement of property $\pmb v_i$. As $f_\phi$ is a random process, the index, $t \in T$, is used to emphasize that the observation $\pmb v_i$ using process $f_{\phi}$ may differ across repeated trials, typically performed sequentially in time. 

In many settings, the measurement process may suffer from known or unknown confounds created in the process of measurement. For example, when taking a magnetic resonance image (MRI)
of a brain, the MRI may be corrupted by motion (movement) or signal intensity artifacts. The observed data, $\pmb w_{it}$, may therefore be unsuitable for direct inference, and instead is pre-processed via the random process $g_\psi: \pmb {\mathcal W} \rightarrow \pmb {\mathcal X}$ to reduce measurement confounds. Here, $\psi \in \pmb \Psi$ characterizes the pre-processing procedure chosen, such as motion or other artifact correction in our MRI example. We define $\pmb x_{it} = g_\psi \circ f_{\phi}(\pmb v_i, t)$ as the pre-processed measurement of $\pmb v_i$ for subject $i$ from measurement index $t$. 
Let $\delta: \pmb{\mathcal X} \times \pmb {\mathcal X} \rightarrow \realn_{\geq 0}$ be a  metric. We use simplified notations such as $\delta_{i,t,t'} = \delta\parens*{\pmb x_{it}, \pmb x_{it'}}$ and $\delta_{i,i', t, t''} = \delta\parens*{\pmb x_{it}, \pmb x_{i't''}}$.

Data repeatability can be considered as a function of the combination of an acquisition procedure, $\phi$, and a chosen pre-processing procedure, $\psi$. Of course, it can be defined exclusively for a subset of the data generating procedure. For instance, when the data has already been collected, the researchers may only be able to manipulate pre-processing, $\psi$, and not acquisition, $\phi$, procedures. Then, one intended use of the repeatability measure is to optimize over those aspects of the measurement process the researcher is able to manipulate: $\psi_* = \argmax_{\psi \in \pmb \Psi} u\parens*{\psi}$, where $u$ is an unspecified repeatability measure. 

Although we will define discriminability with the general framework above, the following additive noise model is a useful special case that maintains tractability: 
\begin{align}
    \pmb x_{it} = \pmb v_i + \pmb \epsilon_{it} \label{eq:additive}
\end{align}
where $\pmb \epsilon_{it} \distas{ind} f_\epsilon$, and $\var{\pmb \epsilon_{it}} < \infty$ with $\expect{\pmb \epsilon_{it}} = \pmb c$. This model is still free of parametric distributional assumptions and contains the MANOVA scenario as a special case. 

\subsection{Definition of Discriminability}
\label{sec:defin_discr}

If the measurement procedure is effective, we would anticipate that our physical property of interest for any subject $i$, $\pmb v_i$, would differ from that of another subject $i'$, $\pmb v_{i'}$. Thus, an intuitive notion of reliability would  expect that subjects would be more similar to themselves
than to other subjects. Specifically, we would expect that $\pmb x_{it}$ is more similar to $\pmb x_{it'}$ (a repeated measurement on subject $i$) than to $\pmb x_{i't''}$ (a measurement on subject $i'$ at time $t''$) for a good measurement. The (population) discriminability is defined as the probabilistic variation of this concept. That is, the probability that  measurements are closer within a subject than between subjects:
\begin{align*}
    D\parens*{\psi, \phi} = \prob{\delta_{i,t,t'} < \delta_{i,i',t,t''}}.
    \label{eqn:discr_true}
\end{align*}
Similar to the fingerprinting index, discriminability is well defined as long as $D(\psi, \phi)$ is equal for all $i, i', t, t', t''$ (such that $i \not = i', t \not = t'$).
That is, this definition assumes that discriminability does not depend on the specific subjects and measurements being considered, a form of exchangeability. The Gaussian (M)ANOVA models in 
(\ref{eq:manova}) are consistent with this form of exchangeability. One could consider a form of population averaged discriminability where $D$ is averaged over subject-specific probabilities; however, we leave this discussion for future work.

To estimate discriminability, assume that for each individual, $i$, we have $s$ repeated measurements. Sample discriminability is then defined as: 
\begin{align}
    \hat{D} = \frac{
    \sum\limits_{i = 1}^n \sum\limits_{t = 1}^s \sum\limits_{t' \neq t}
    \sum\limits_{i' \neq i} \sum\limits_{t'' = 1}^s \indicator{\delta_{i, t, t'} < \delta_{i, i', t, t''}}
    }{
    n \cdot s \cdot (s-1) \cdot (n-1) \cdot s
    }.
\end{align}
Under this definition, $\hat D$ represents the fraction of observations where $\pmb x_{it}$ is more similar to $\pmb x_{it'}$ than to the measurement $\pmb x_{i't''}$ of another subject $i'$, for all pair of subjects $i \not= i'$ and all pairs of time points $t \not= t'$.  It is perhaps useful to let $D^* = 2 D - 1$  to transform discriminability to range between $0$ to $1$, similar to ICC. However, we do not follow that convention here to maintain the direct relationship 
between the estimator and the population quantity of interest. As specified the
sample discriminability estimator is both unbiased and consistent for the population discriminability under the additive noise model, \ref{eq:additive} (see the proof in Appendix \ref{appendix:discrim}). 

\subsection{Relationships Between Discriminability and Intraclass Correlation Coefficients}
\label{sec:anova_approx}
\begin{figure}
    \centering
\includegraphics[scale = .6]{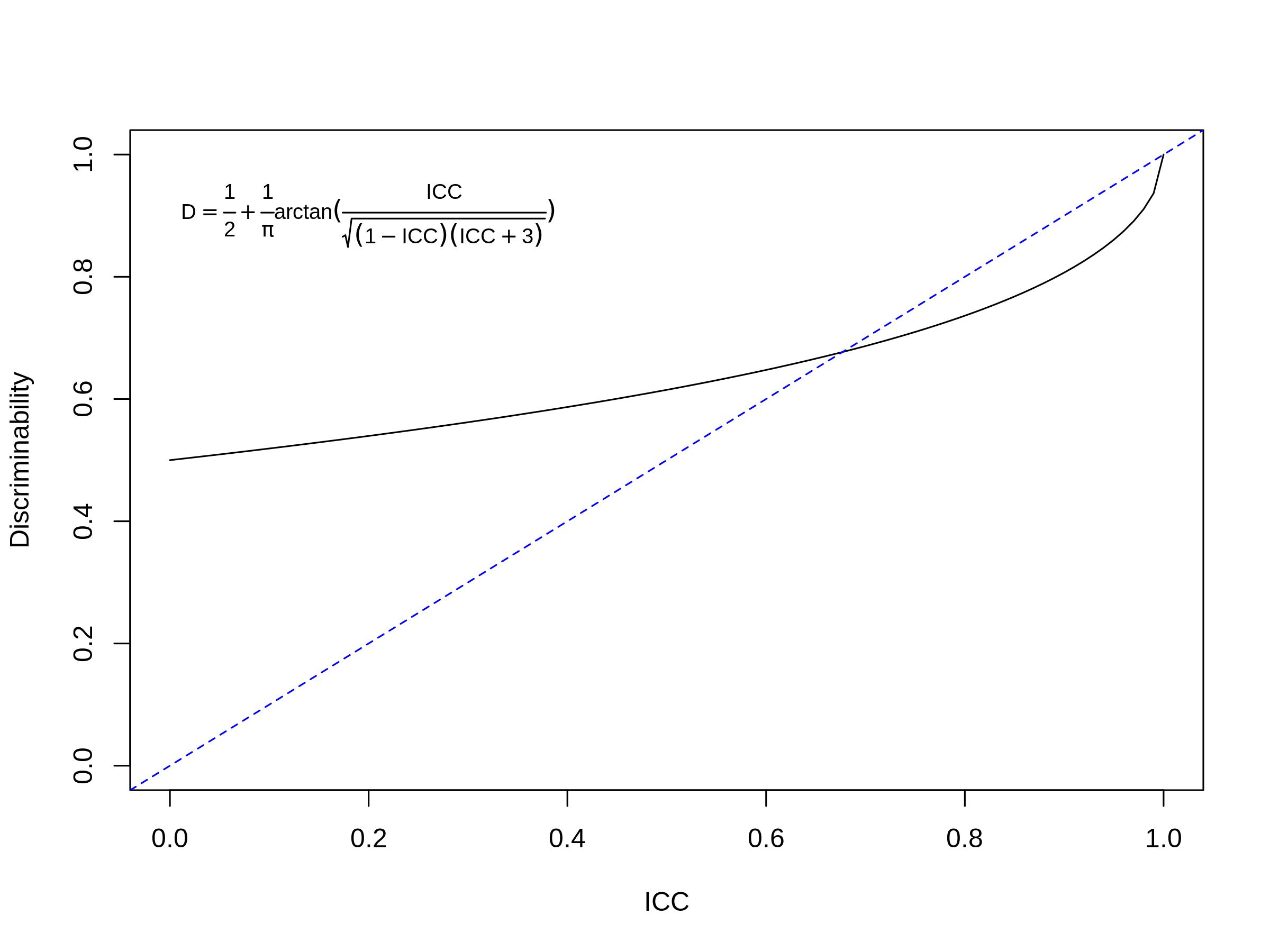}
\caption{The relation between discriminability and ICC under the ANOVA model with Gaussian random effects. See Section \ref{sec:anova_approx}. }
    \label{fig:d_icc}
\end{figure}

Interestingly, under the ANOVA model in (\ref{eq:manova}) with $l=1$, discriminability 
is deterministically linked to ICC. It is relatively easy to argue and
instructive on the relationship between these constructs, and therefore
we present the argument here. Considering a Euclidean distance as the metric, discriminability ($D$) is: 
\begin{align*}
    D & = \prob{\abs*{x_{it} - x_{it'}} < \abs*{x_{it} - x_{i't''}}} \\
    & = \prob{\abs*{e_{it} - e_{it'}} < \abs*{\mu_i - \mu_{i'} + e_{it} - e_{it''}}}
    \\ & \stackrel{def}{=} \prob{|A| < |B|}
\end{align*}
for $i \not= i'$, $t \not = t'$. Then, $(A, B)^t$ follows a joint normal distribution, with mean vector $\mathbf{0}$ and covariance matrix 
$\left( \begin{smallmatrix} 2\sigma^2&\sigma^2\\ \sigma^2&2\sigma_\mu^2 + 2\sigma^2 \end{smallmatrix} \right)$. Hence:
\begin{align*}
    D & = 1 - \frac{\arctan \left(\frac{\sqrt{\sigma^2 (3\sigma^2 + 4\sigma^2_\mu)}}{\sigma_\mu^2} \right)}{\pi}
     = \frac{1}{2} + \frac{1}{\pi} \arctan \left( \frac{\mathrm{ICC}}{\sqrt{ (1 - \mathrm{ICC})(\mathrm{ICC} + 3)} } \right)
    .
\end{align*}
Therefore, in this setting {\it discriminability and ICC are deterministically linked with a non-decreasing transformation under the ANOVA model with Gaussian random effects}. Figure \ref{fig:d_icc} shows a plot of the non-linear relationship. For
an ICC of roughly 0.68, the two measures are equal, with discriminability
being smaller for ICCs larger than 0.68 and larger for ICCs lower.

Recall, the optimal correlation between two univariate measurements equals to a non-decreasing function of the ICC of each of the measurement. Since discriminability is deterministically linked to ICC via a strictly increasing function, this property also holds for discriminability.  Another scenario where the repeatability measure may become critical is in prediction with multivariate predictors. The optimal prediction error in terms of the Bayes error rate of a classification task can be bounded by a decreasing function of discriminability of the multivariate predictors \citep{bridgeford2021eliminating}. 
Therefore, optimizing discriminability minimizes an upper bound on the irreducible error due to data generating process. In general, we expect improved data repeatability leads to improved optimal performance of downstream prediction tasks, which holds true for discriminability nonparametrically.
In addition, it is interesting to note that ICC inherits this property exactly under Gaussian assumptions, as it holds for any one-to-one transformation of discriminability.


For multivariate measurements, the relationship gets more complex. Under Model \ref{eq:manova} with Euclidean distances, discriminability becomes: 
\begin{align*}
D & = P(||\pmb x_{it} - \pmb x_{it'} || - ||\pmb x_{it} - \pmb x_{i't''}|| < 0) \\
& = P(||\pmb e_{it} - \pmb e_{it'} || - ||\pmb e_{it} - \pmb e_{i't''} + \pmb\mu_i - \pmb\mu_{i'}|| < 0) \\
& \stackrel{def}{=} P(||\pmb A|| - ||\pmb B|| < 0), 
\end{align*}
where $A$ and $B$ are jointly multivariate normal with means 0,
variances $2\pmb \Sigma$ and $2\pmb \Sigma + 2\pmb \Sigma_{\pmb \mu}$, respectively,
and covariance, $\pmb \Sigma$.
Note that $Z \stackrel{def}{=} \pmb A^t \pmb A - \pmb B^t \pmb B $
is an indefinite quadratic form of the vector $(\pmb A^t, \pmb B^t)^t$
(around a matrix whose block diagonal entries are an identity matrix and the negative of an identity matrix). Thus, $Z$ can be decomposed as a linear combination of independent one degree of freedom $\chi^2$ variables \citep{provost1996exact}: 
\begin{align}
\label{eq:decomposition}
Z & \stackrel{d}{=} \sum_{u = 1}^r \lambda_u U_u
- \sum_{u = r + 1}^{r + w} \lambda_u' U_u
,
\end{align}
where $\lambda_1, \dots, \lambda_r$ are the positive eigenvalues of 
$
\left (
\begin{array}{cc}
2\pmb \Sigma & -\pmb \Sigma \\ 
\pmb \Sigma & -2\pmb \Sigma - 2\pmb \Sigma_{\pmb \mu}
\end{array}
\right )
$, 
$\lambda_{r + 1}', \dots, \lambda_{r + w}'$ are the absolute values of the negative eigenvalues of 
$
\left (
\begin{array}{cc}
2\pmb \Sigma & -\pmb \Sigma \\ 
\pmb \Sigma & -2\pmb \Sigma - 2\pmb \Sigma_{\pmb \mu}
\end{array}
\right )
$, 
$U_1, \dots, U_{r + w}$ are IID $\chi^2$ variables. 

\begin{figure}
    \centering
\includegraphics[scale = .6]{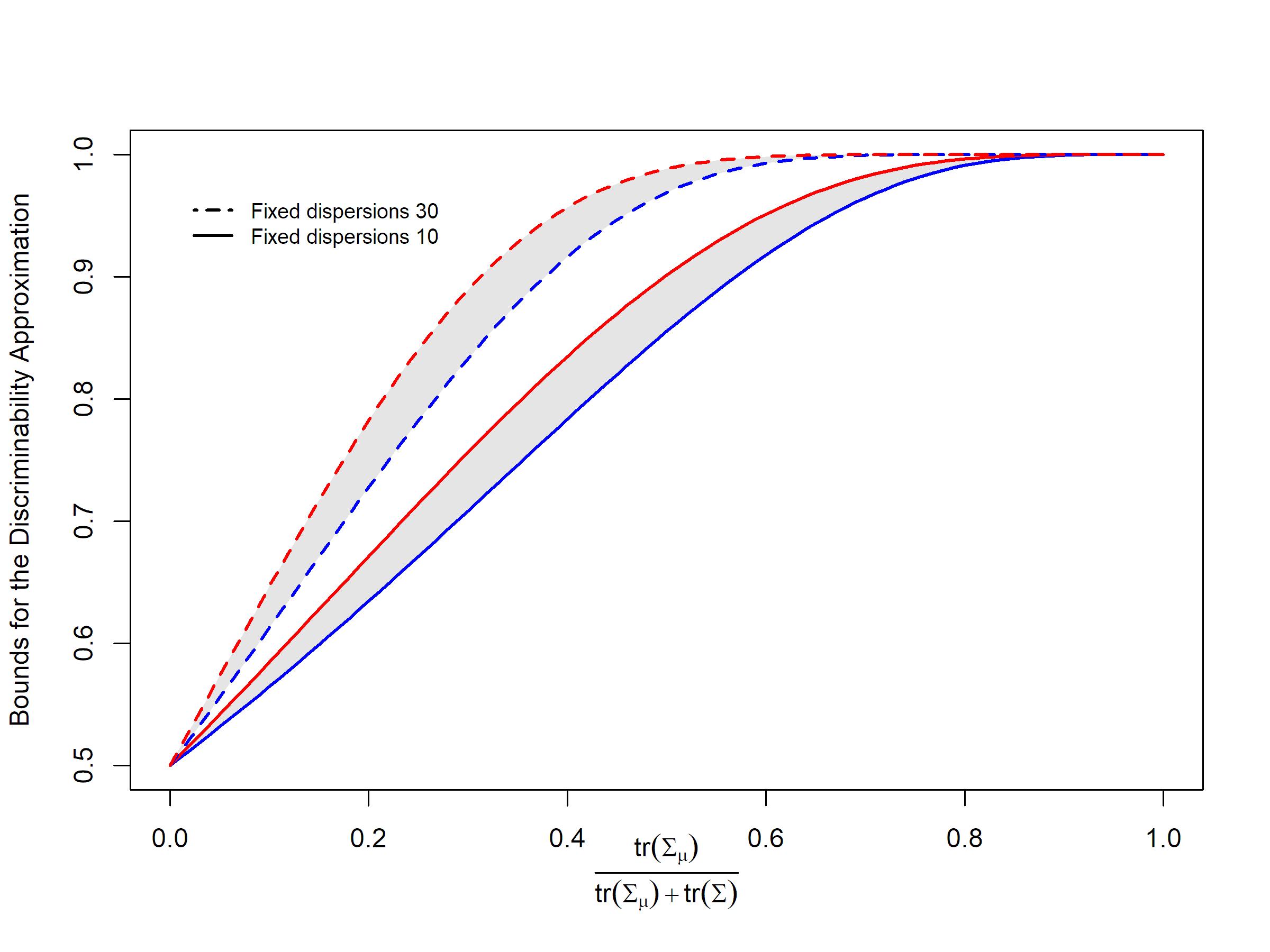}
\caption{Non-decreasing bounds of the discriminability approximation (\ref{eq:approxF}) using functions of I2C2 under the MANOVA model with random Gaussian effects. The dispersion measures, defined as $V_1^2/W_1$ and $V_2^2/W_2$, are fixed at $10$ or $30$. The upper (red) and lower (blue) bounds are color coded, respectively. The dispersion $10$ scenario is plotted with solid lines whereas the dispersion $30$ scenario is plotted with dashed lines. $V_1, W_1$ (or $V_2, W_2$) are the sum and the sum of squares of the positive (or negative) eigenvalues from the distributional decomposition (\ref{eq:decomposition}). See Section 
\ref{sec:anova_approx}
. } 
    \label{fig:bound_dispersion}
\end{figure}

Although this does not result in a deterministic link between $D$ and I2C2, it can be shown that there exist approximations matching the first two moments of $\sum_{u = 1}^r \lambda_u U_u$ and $\sum_{u = r + 1}^{r + w} \lambda_u' U_u$. Furthermore, the approximation of $D$ can be bounded by two non-decreasing functions of I2C2 (Appendix \ref{appendix:mono}). Specifically, the resulting discriminability approximation has the form of a CDF value of an F-distribution, 
\begin{align*}
D = P(Z \leq 0)
& \approx F_{F\left(\frac{V_1^2}{W_1}, \frac{V_2^2}{W_2}\right)}\left(\frac{V_2}{V_1}\right), 
\label{eq:approxF}
\end{align*}
where $V_1, W_1$ (or $V_2, W_2$) are the sum and the sum of squares of the absolute values of the positive (or negative) eigenvalues. Moreover, when $V_1, W_1, V_2, W_2$ are constant, the approximation is bounded by a non-decreasing interval of I2C2 (Figure \ref{fig:bound_dispersion}): 
\begin{align*}
 F_{F\left(\frac{V_1^2}{W_1}, \frac{V_2^2}{W_2}\right)}\left\{
 f_1(\Lambda_{tr})
 \right\}
\leq 
    F_{F\left(\frac{V_1^2}{W_1}, \frac{V_2^2}{W_2}\right)}\left(\frac{V_2}{V_1}\right)
    \leq 
    F_{F\left\{\frac{V_1^2}{W_1}, \frac{V_2^2}{W_2}\right\}}\left\{
    f_2(\Lambda_{tr})
    \right\}, 
\end{align*}
where $f_1(v) = 1 + v/(1-v)$ and $f_2(v) = 1 + (4/3) \cdot v/(1-v)$ are both non-decreasing functions. 

\subsection{Relation with Fingerprinting and Rank Sums}
\label{sec:rs}
In the test-retest setting where the fingerprint index is defined under a nonparametric model, such as (\ref{eq:additive}),
it can be shown that the fingerprint index in general converges to $0$ as $n$ increases (Appendix \ref{appendix:discrim_fp}). 
Even in datasets with a fixed sample size of $n$ and an identical discriminability, $D$, the values of the fingerprinting index may vary depending on the underlying data generating processes. 
This highlights a fact that we will return to in the simulations: the fingerprint index is not invariant for smaller sample sizes (below 10 to 15, say), even when discriminability is held constant. 

The relationship between population discriminability and the fingerprint index is dependent on the specific distribution of the data,  
and there is no direct relationship in their estimators. However, interestingly, the sample discriminability can be rewritten as a function of a form of rank sums. In addition, the specific form of rank sum statistics, $R_n$, can be transformed to a consistent estimator of discriminability. These highlights that the sample discriminability estimator retains rank information that the fingerprint statistic discards. 

To elaborate, denote the $n$ by $n$ inter-measurement distance sub-matrix as $\pmb D^{t, t'} = \left( \delta_{i, i', t, t'} \right)_{i, i' = 1, \dots, n}$. Let the combined $n\cdot s$ by $n\cdot s$ distance matrix be $\pmb D = \left(\pmb D^{t, t'} \right)_{t, t' = 1, \dots, s}$, which consists of $s$ by $s$ blocks where the $(t, t')$ block is $\pmb D^{t, t'}$. Let $r_{i, i'}^{t, t'}$ denote the ranking within rows in the combined distance matrix $\pmb D = \left(\delta_{i,i', t, t'} \right)$. Throughout, we assign the maximum ranks for ties.  It can then be shown (Appendiex \ref{appendix:alter}) that a consistent estimator of discriminability using only the ranks is: 
\begin{equation}
\label{eq:est_new1}
\Tilde{D} = \frac{n^2s^2(s-1) - \sum_{t=1}^s \sum_{t'\not = t} \sum_{i=1}^n r_{ii}^{tt'} }{ns(s-1)(n-1)s}. 
\end{equation}

In fact, the standard rank sum statistic, (\ref{eq:rs}), can be transformed to an estimate of discriminability.  In a test-retest setting with $s = 2$, instead of ranking the combined distance matrix, $\pmb D$, let $r_{ij}$ be the rank of $\delta_{i,j}^{1,2}$ among $\delta_{i,1}^{1,2}, \dots, \delta_{i,n}^{1,2}$, which ranks the row of the inter-measurement distance sub-matrix $\pmb D^{1,2}$. This transformation of the rank sum statistic, $R_n$, forms an unbiased and consistent estimator of $D$: 
\begin{equation}
\label{eq:est_RS}
\hat{D}_{rs} = \frac{\sum_{i=1}^n (n - r_{ii})}{ n(n-1)} = \frac{n^2 - R_n}{n(n-1)}. 
\end{equation}

If there exist multiple measurements for each subject, for all the pairs of distinct $t_1$ and $t_2$, the rank sum statistic and estimation can be calculated between the $t_1$-th measurements and the $t_2$-th measurement. 
Comparing to $\hat{D}$ and $\Tilde{D}$, the rank sum statistic does not involve any ranking information from the diagonal blocks in the combined distance matrix,
$\pmb D^{t,t}, t = 1, \dots, s$,. This may result in a larger standard error for estimation and a lower power for inference using the rank sums. However, it provides some robustness against mean shift batch effects, as demonstrated in Section \ref{sec:batch}.

\section{Permutation Tests to Confirm Existence of Data Repeatability} \label{sec:test}

The repeatability of the process that generates data $\pmb x_{it}$ can be characterized by its dependence on a invariant subject-specific true signal $\pmb v_i$, which repeatedly enters the data generating process of subject $i$. To design a statistical test that confirms existence of data repeatability, the null hypothesis can be $H_0: \pmb x_{it} \indep \pmb v_i$, that is, the independence between measurements and the unobserved subject-specific invariant true signal.

To enhance robustness across different data distributions, permutation tests with few parametric assumptions are preferred in this setting. 
Under Model (\ref{eq:additive}), $H_0: \pmb x_{it} \indep \pmb v_i$ implies $\pmb x_{it} = \pmb \epsilon_{it}$, which guarantees exchangeability of any repeatability statistics defined in the previous sections. 
Specifically, the null distribution of a repeatability statistic $T(X_{it}: i, t)$ defined in previous sections is invariant against permutations of the subject labels. For example, for measurements with multiple time points, the subject labels can be permuted within each of the time points, so that under $H_0$ we have $T(X_{it}: i, t) \overset{d}{=} T(X_{\pi_{it},t}: i, t)$ for any set of permutations $\{(\pi_{1t}, \dots, \pi_{nt}): t\}$ of $(1, \dots, n)$. 
Therefore, permutation tests against the weaker null of exchangeability can be conducted. 
On the other hand, the alternative hypothesis, $\pmb x_{it} \not\indep \pmb v_i$, provides evidence against no repeatability in which case repeated measurements reveal information on unobserved subject-specific effects.
In summary, we have
\begin{align*}
    & H_0: \pmb x_{it} \indep \pmb v_i \\
    & H_A: \pmb x_{it} \not\indep \pmb v_i. 
\end{align*}
However, the performance of these repeatability statistics in the permutation tests under different model settings is less mathematically clear.  
In Section \ref{sec:numerical} we present numerical results, including the consequences of deviations from the ANOVA model. 



In practice, we note that under the null $H_0$ and the implied exchangeability, the repeatability statistics described in Section \ref{sec:repeatability} have an invariant distribution under permutation of the observed samples via Monte Carlo. 
In fact, Monte Carlo resampling \citep{good2013permutation} eliminates the typically impossible task of looping over the space of permutations, a set which can have a cardinality up to $(n!)^s$ ($n$ subjects each measured on $s$ occasions).  A P-value approximation is obtained by calculating the proportion of permutations where the simulated value of the repeatability statistic is more extreme than the observed. 




\section{Numerical Experiments} \label{sec:numerical}

\subsection{Univariate ANOVA Simulations} \label{sec:anova}

We first evaluate the estimation and testing powers under the ANOVA model (\ref{eq:manova}) with $l=1$ or when its Gaussian assumptions are violated. $t \in \{1, 2\}$, $\sigma^2 = 5, \sigma_\mu^2 = 3$. The number of subjects, $n$, ranges from $5$ to $40$.  In addition to the correct Gaussian model, consider the following lognormal misspecification: 
\begin{align*}
\mu_i & \distas{d} \Lognorm{0, \sigma_\mu^2};\; \log(\mu_i) \distas{d} \Norm{0, \sigma_\mu^2},\\
e_{it} & \distas{d} \Lognorm{0, \sigma^2};\; \log (e_{it}) \distas{d} \Norm{0, \sigma^2},
    \label{eq:lognormal_anova}
\end{align*}
where we still define $ICC = \textrm{var}(\mu_i) / (\textrm{var}(\mu_i) + \textrm{var}(e_{it}))$, but now $\textrm{var}(e_{it}) = (\exp(\sigma^2) - 1) \cdot \exp(\sigma^2)$. Note that the relation between discriminability and ICC does not hold in this setting. 

For $1$,$000$ iterations, estimates of discriminability (using $\Tilde{D}$ in the Equation \ref{eq:est_new1}), the rank sum estimator  ($\hat{D}_{rs}$ in Equation \ref{eq:est_RS}),  estimations of ICC using one-way ANOVA, 
estimations of the fingerprint index (using $\hat F_{index}$ in Section \ref{sec:fpt}) were recorded and compared to their theoretical true values (for discriminability and ICC) or its simulated average value (for the fingerprint index, with $10$,$000$ simulations).  Within each iteration, we also conducted permutation tests against exchangeability, each with $1,000$ Monte Carlo simulations, using the same estimators. F-tests using the ICC  F-statistics were also conducted. The proportion of rejections (power curves) by iterations were plotted. 

\begin{figure}
    \centering
\includegraphics[scale = .72]{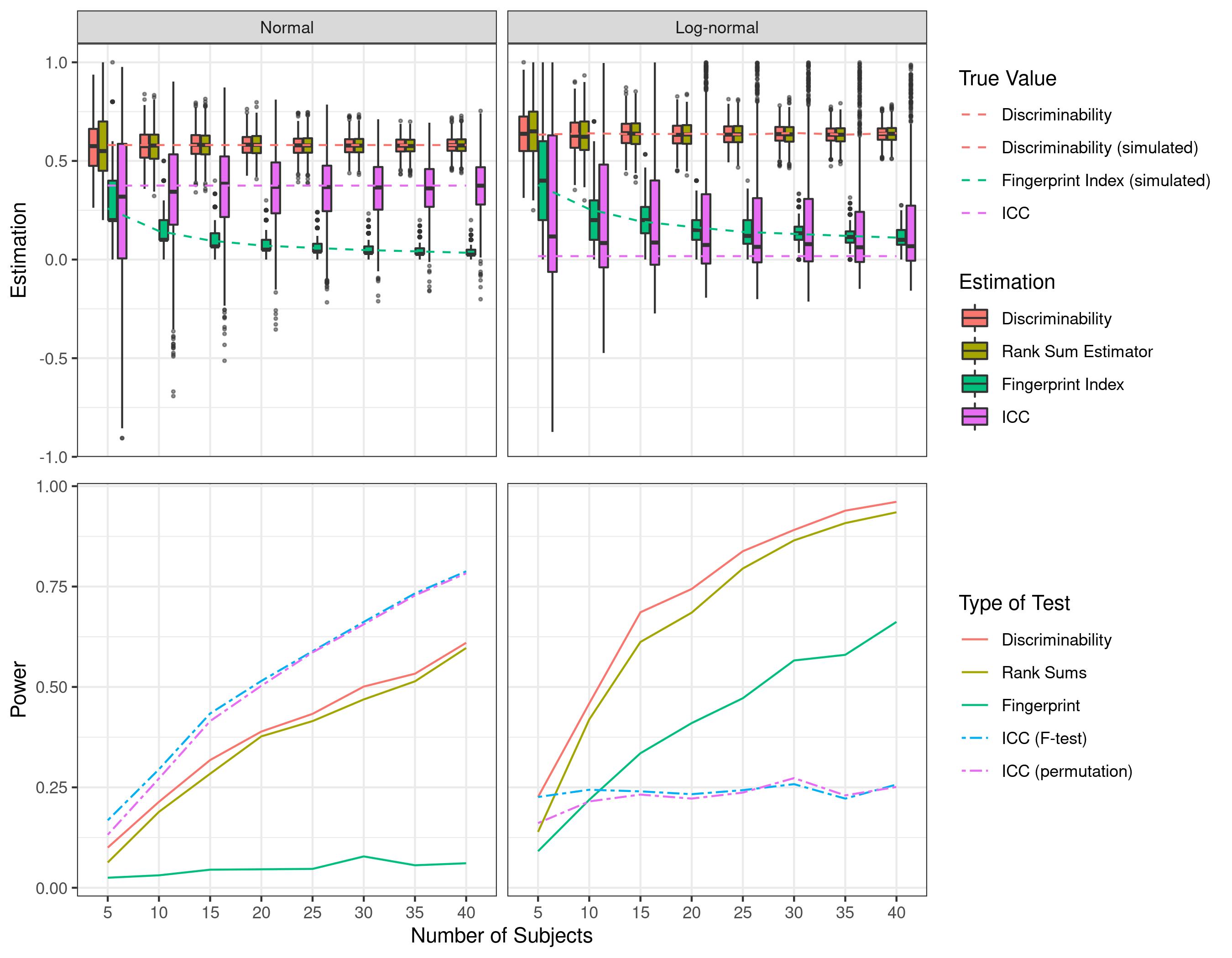}

\caption{ANOVA simulations when the Gaussian assumption is satisfied (left) or violated with logarithm transformations (right). Simulated distributions of estimators are plotted on the top, including the discriminability estimation (using the estimator $\Tilde{D}$ or the rank sum version $\hat{D}_{rs}$), the fingerprint index estimation, and the ICC estimation. Simulated permutation test powers are plotted on the bottom, where solid lines and dotted lines represent nonparametric and parametric statistics, respectively. $\sigma^2 = 5$. $\sigma_\mu^2 = 3$. $n$ ranges from $5$ to $40$. $1$,$000$ iterations in total. See Section \ref{sec:anova}. 
}
    \label{fig:anova}
\end{figure}

As expected, when the parametric assumption is satisfied, all estimators are distributed around their true values (Figure \ref{fig:anova}). However, the distribution of the fingerprint index is skewed. In addition, a higher fingerprinting index estimation with fewer subjects does not imply better repeatability, compared the lower estimation with more subjects. Of note, the true ICC and discriminability remain constant as the sample size increases. Thus, this reinforces the theoretical intuition that the fingerprint index is not directly comparable across sample sizes. In terms of hypothesis testing power, tests using statistics associated with the ICC produce higher power, if the Gaussian model is correctly specified. The discriminability estimator using the whole combined ranking matrix shows slight advantage in power compared to the rank sum estimator, which only uses rank sums within a submatrix of the combined distance matrix. Lastly, switching to fingerprinting results in a loss in testing power. 

We repeated the simulation in an otherwise similar setting where normality does not hold: $\textrm{var}(\mu_i) = (\exp(\sigma_\mu^2) - 1) \cdot \exp(\sigma_\mu^2) \approx 383$, $\textrm{var}(e_{ij}) = (\exp(\sigma^2) - 1) \cdot \exp(\sigma^2) \approx 21878$, and $ICC$ is around $0.017$. Because of the model misspecification, ICC is overestimated with relatively large variation. As for hypothesis testing power, the discriminability estimator, rank sum and the fingerprint index estimator outperform their parametric counterparts (likely due to their nonparametric foundations). Discriminiability, $\Tilde{D}$, again has higher power than $\hat{D}_{rs}$, likely as a byproduct of including more ranking information. The $\hat{F}_{index}$ statistic has a loss in power over disciminability or rank sums, but now outperforms the estimators based on the parametric assumptions of ICC or F-statistics. 

\subsection{MANOVA Simulations} \label{sec:manova}

Next, we consider the MANOVA model (\ref{eq:manova}) and a similar misspecification with element-wise log-transformations on the subject mean vectors, $\pmb \mu_i$, and the noise vectors, $\pmb e_{it}$. $t = 1, 2$. $n$ ranges from $5$ to $40$.  We simulated data with
$\pmb \Sigma = \sigma^2 \pmb Q$, 
$\pmb \Sigma_{\pmb\mu} = \sigma_\mu^2 \pmb Q$, 
and $\pmb Q = \pmb I (1 - \rho) + \pmb 1 \pmb 1 ^ t \rho$ (an $l\times l$ exchangeable correlation matrix, with off diagonals $\rho$). 
Let $\sigma^2 = 5, \sigma_\mu^2 = 3, \rho = 0.5, l = 10$. For $1$,$000$ iterations, the estimations and the permutation test (each performed with $1,000$ Monte Carlo simulations) power were compared for discriminability, the rank sum estimator of discriminability, the fingerprint index, the sample ICC, $\hat{\lambda}$, calculated with the first principal components from the measurements, and I2C2. 

\begin{figure}
    \centering
\includegraphics[scale = .72]{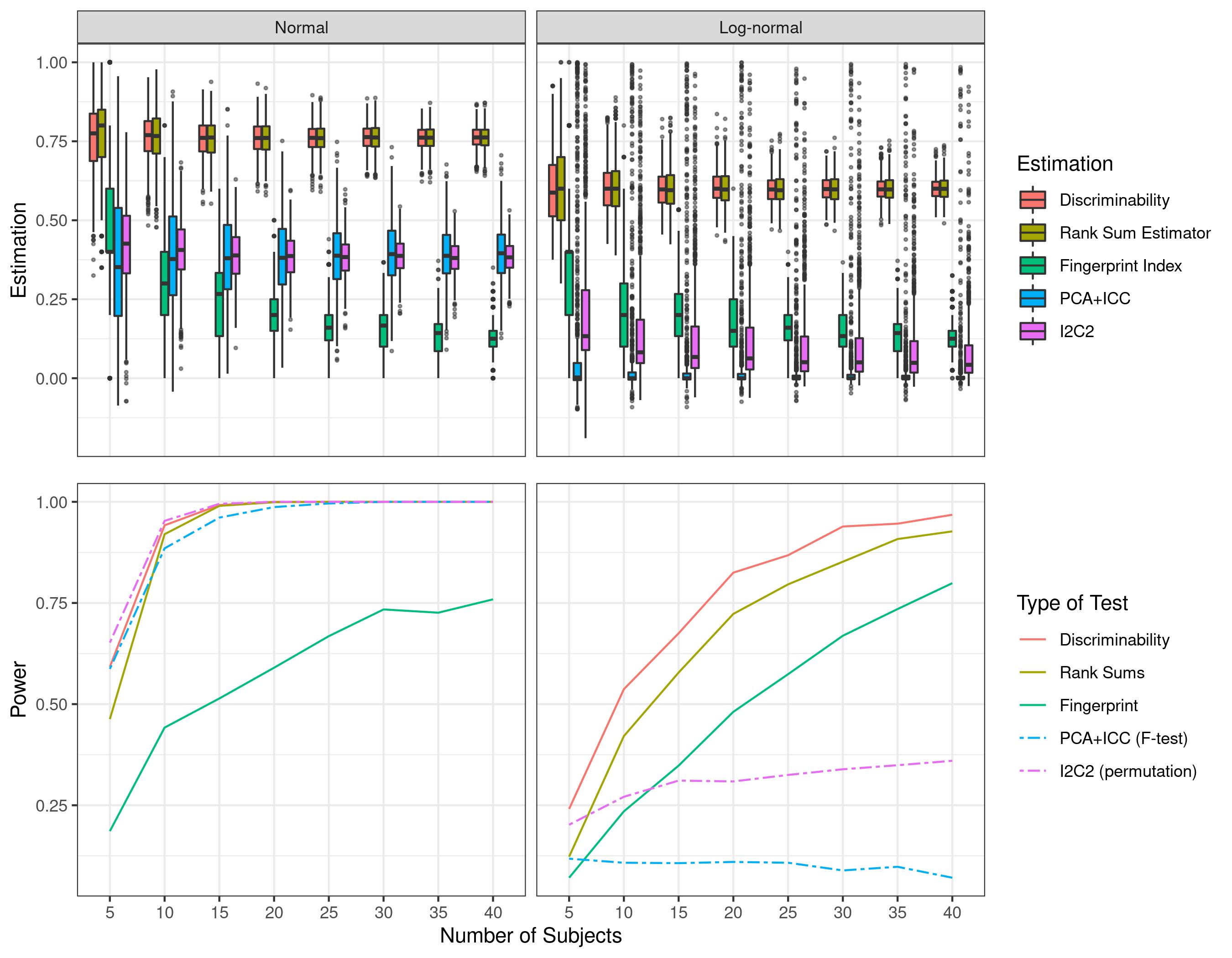}

\caption{MANOVA simulations when the Gaussian assumption is satisfied (left) or violated with element-wise logarithm transformations (right). Simulated distributions of estimators are plotted on the top, including the discriminability estimation (using the estimator $\Tilde{D}$ or the rank sum version $\hat{D}_{rs}$), the fingerprint index, and the I2C2. Simulated permutation test powers are plotted on the bottom, where solid lines and dotted lines represent nonparametric and parametric statistics, respectively. $\sigma^2 = 5$. $\sigma_\mu^2 = 3, \rho = 0.5, l = 10$. $n$ ranges from $5$ to $40$. $1$,$000$ iterations in total. See Section \ref{sec:manova}. 
}
    \label{fig:manova}
\end{figure}

When the Gaussian assumption is satisfied, I2C2 outperforms other statistics, and most statistics produce higher testing power compared to the fingerprint index (by a large margin Figure \ref{fig:manova}). Note that the strategy of conducting PCA before ICC also shows advantage over discriminability in power when the sample size is as small as $5$, but power converges with larger sample sizes. 
When normality is violated, the nonparametric statistics (discriminability, rank sums, and fingerprinting outperform the parametric methods in power with any sample sizes greater than $10$. The discriminability estimator provides the best power under the multivariate lognormal assumptions.

\subsection{Batch Effects} \label{sec:batch}

Consider the ANOVA model (\ref{eq:manova}) with $l=1$ where each subject is remeasured for $s$ times, $s > 2$. We evaluate two types of batch effects, mean shifts and scaling factors \citep{johnson2007adjusting}.  For the mean shifts, we replace the subject means, $\mu_i$'s, with the batch specific means $\mu_{it}$'s defined as: 
\begin{align*}
    \mu_{i1} & \distas{d} \Norm{0, \sigma_\mu^2} \\
    \mu_{it} & = \mu_{i1} + t, t = 2, \dots, s.
\end{align*}
Without loss of generality, consider the first batch as a reference batch, where $\mu_{i1}$'s follow the same distribution as the previous $\mu_i$'s. For the $t$-th batch, there exists a mean shift, $t$, from the reference batch for all subjects. 
The scaling effects were applied on the noise variances as:
\begin{align*}
    e_{i1} & \distas{d} \Norm{0, \sigma^2} \\
e_{it} & \distas{d} \Norm{0,t\sigma^2}, t = 2, \dots, s. 
\end{align*}

Note that by default $\hat{D}_{rs}$ does not handle multiple repeated measurements. In order to thoroughly compare the original discriminability estimator, $\Tilde{D}$ (\ref{eq:est_new1}), and the rank sum based estimator, $\hat{D}_{rs}$ (\ref{eq:est_RS}), at each time point $t = 2, \dots, s$, we considered the following different repeatability estimators. First, we considered only the first and the $t$-th batches (\textit{first-last}) and computed $\Tilde{D}$ and $\hat{D}_{rs}$ directly. Secondly, we used all measurements up to the $t$-th time point (\textit{all batches}). The estimator $\Tilde{D}$ can be directly calculated, whereas the $\hat{D}_{rs}$ can be generalized by averaging on all pairs of time points. Lastly, we considered a special case where we averaged over only the pairs of time points between the first and the rest (\textit{first-rest}) for both $\Tilde{D}$ and $\hat{D}_{rs}$. In total, six multi-time-point discriminability estimators were considered, where three of them are $\Tilde{D}$-based and the other three are $\hat{D}_{rs}$-based.

We simulated $s = 15$ batches in total with $\sigma^2 = 3, \sigma_\mu^2 = 5$ and let the number of subjects, $n$, range from $5$ to $40$. For $1$,$000$ iterations, the estimations and the permutation test (each with $1,000$ Monte Carlo iterations) power of the six estimators described above are plotted.

\begin{figure}
    \centering
\includegraphics[scale = .37]{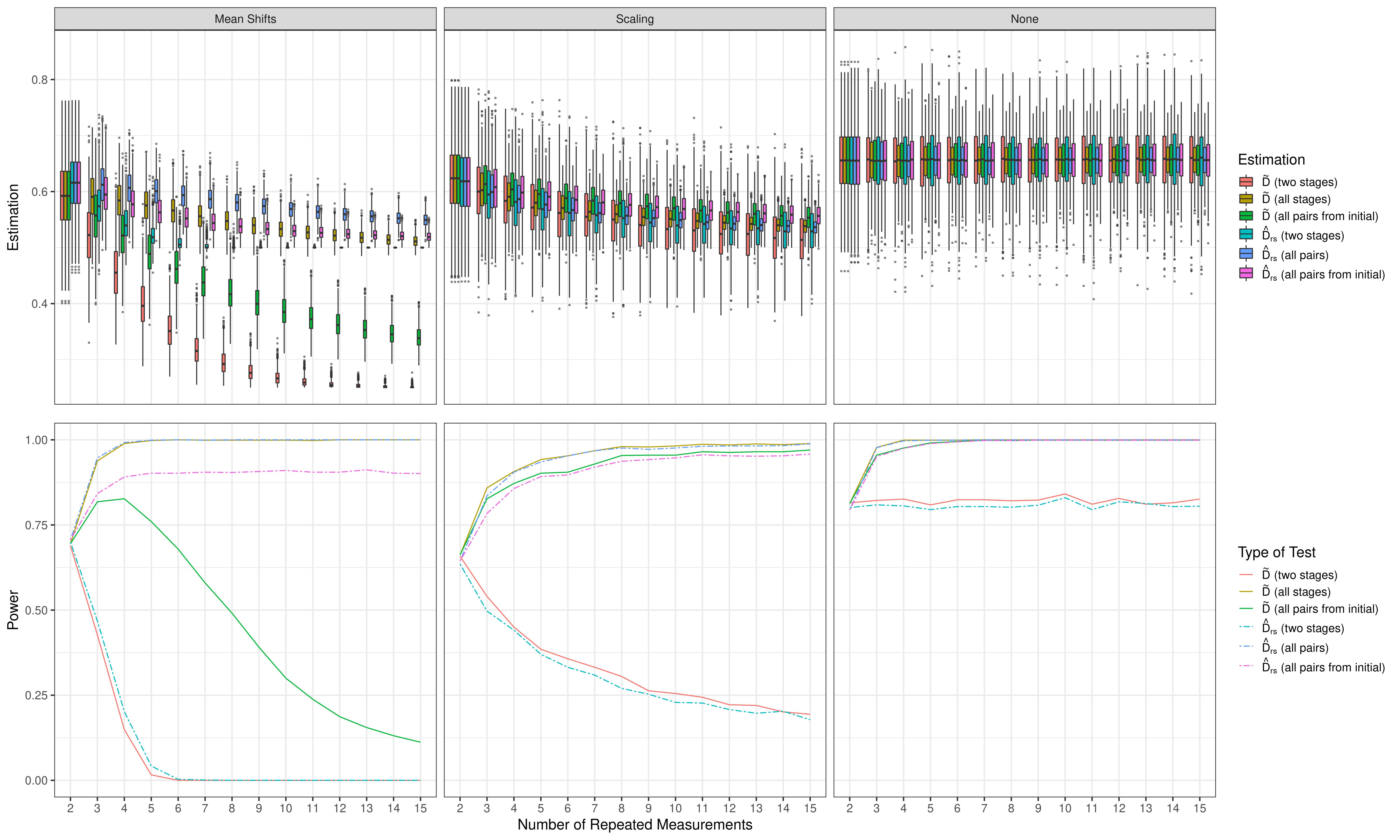}

\caption{Simulations for mean shifts (left), scaling (middle), and no batch effect (right). On the top are simulated distributions of the six discriminability estimators for multiple repeated measurements (Section \ref{sec:batch}). On the bottom are simulated testing powers, where the three $\Tilde{D}$-based (the original discriminability estimator \ref{eq:est_new1}) estimators are solid lines and the three $\hat{D}_{rs}$-based (the rank sum based estimator \ref{eq:est_RS}) estimators are dotted lines. As the number of repeated measurements increased, the estimation was conducted by: taking the first and the last batches (\textit{first-last}), using all batches (\textit{all batches}; averaging over all pairs of batches for $\hat{D}_{rs}$), or averaging only pairs between the first and the rest of the batches (\textit{first-rest}). Different strategies are for thorough comparisons between $\Tilde{D}$ and $\hat{D}_{rs}$, as by default $\hat{D}_{rs}$ only handles two batches. 
$1$,$000$ iterations in total. 
}
    \label{fig:batch}
\end{figure}

For the mean shift only batch effects, the rank sum estimator outperforms discriminability in power with the highest power achieved using all time point pairs (Figure \ref{fig:batch}). The estimation from rank sums is also closer to the batch-effect-free true discriminability, $0.625$. The rank sum method may benefit from the fact that, whenever $t = t''$, it avoids averaging over indicators 
\begin{align*}
    \indicator{\delta_{i, t, t'} < \delta_{i, i', t, t''}} = \indicator{|(t - t') + (e_{it} - e_{it'})| < |(\mu_{it} - \mu_{i't}) + (e_{it} - e_{i't})|}, 
\end{align*}
where the batch difference, $(t - t')$, if larger enough, may force the indicator to be $0$ with high probability, regardless of the true batch-effect-free discriminability level. For example, for the \textit{all pairs from initial} scenario, rank sums outperform discriminability by a huge margin, since batch differences become larger when later batches are compared to the reference batch. 

For the scaling only batch effects, discriminability now outperforms rank sums, regardless of the strategy used. (Using all time points produces the highest power.) This is similar to the case with no batch effects, where having more repeated measurements increases testing power, and the advantage of discriminability over rank sums and the advantage of using all time points are attained.

\begin{figure}
    \centering
    \includegraphics[scale = 0.8]{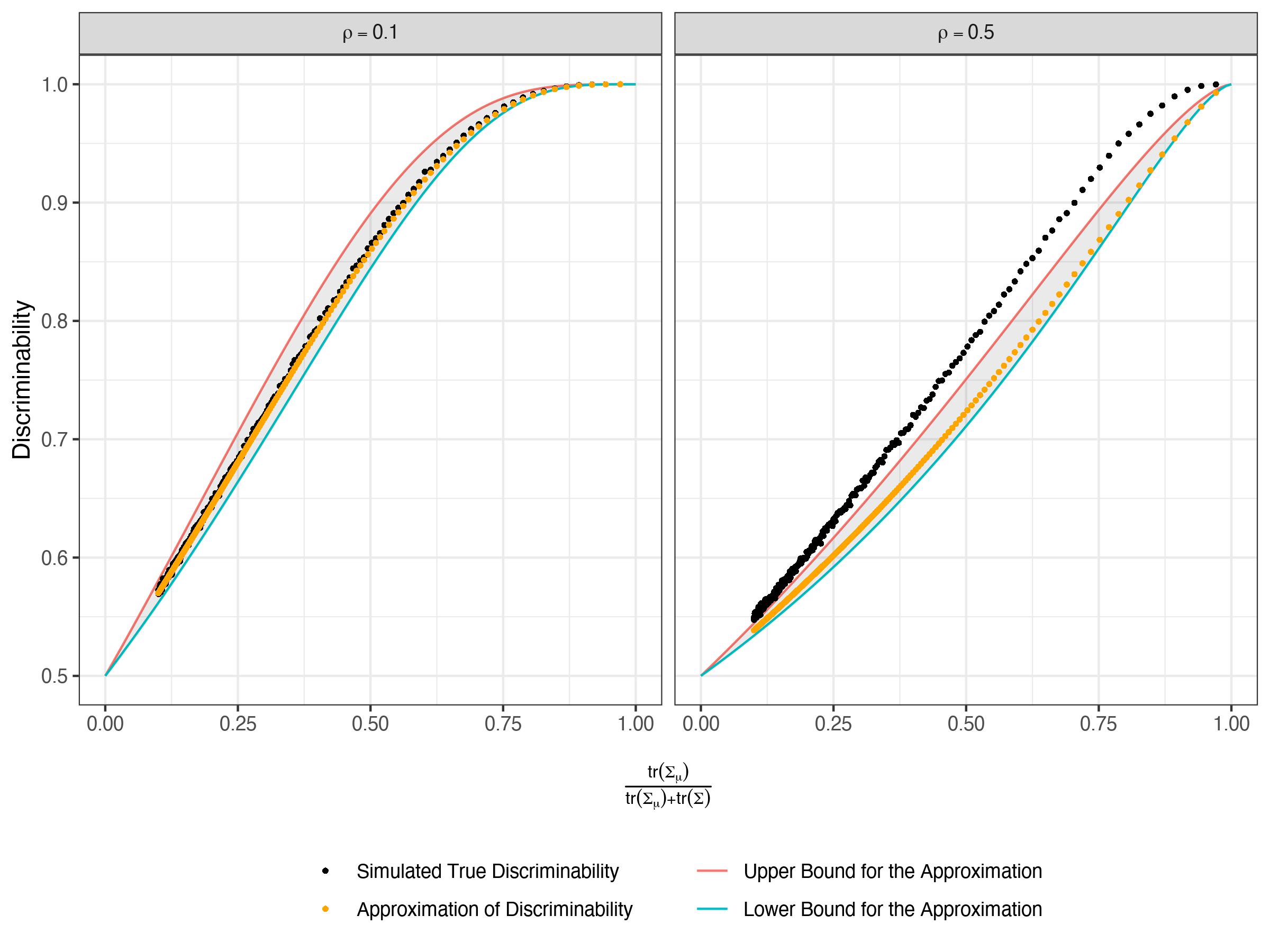}
    
    \caption{Relation between discriminability and I2C2 with smaller ($\rho = 0.1$, left) or larger ($\rho = 0.5$, right) within measurement correlation. The Gaussian MANOVA model in Section \ref{sec:manova} is assumed with $l = 10, n = 20, s = 2$. Covariance matrices, $\pmb \Sigma$ and $\pmb \Sigma_\mu$, are proportional to a matrix with diagonals being $1$ and off-diagonals being $\rho$. Small circles are the simulated ($1$,$000$ iterations) true discriminability with $\sigma_\mu^2 = 100$ and $\sigma^2$ ranging from $3$ to $300$. This shows error of the approximation (\ref{eq:approxF}) is within $0.1$ and the non-decreasing relation holds true even with larger $\rho$ value. 
    } 
    \label{fig:multiapprox}
\end{figure}

\pagebreak

\section{Discussion}

The key contribution of this manuscript is the explicit statement of population quantities
associated with repeatability metrics, the models that support them and elucidating the relationships between them, particularly, the relationship between discriminability, ICC and I2C2. Note the latter is distinct from the known non-decreasing relation between ICC estimates and the F statistic, which guarantees the same ordering and power for the associated permutation test. The fact that ICC and I2C2 have higher power when parametric assumptions are satisfied hints at the potential for improving discriminability estimation. Another potential improvement is the approximation (\ref{eq:approxF}) of the weighted sum of $\chi^2$'s, as it tends to underestimate more $D$ with increasing intra-measurement correlations (Figure \ref{fig:multiapprox}). However, even with the current approximation, the error is within $0.1$ and the non-decreasing relation holds true in the simulations with larger $\rho$ values. 
Further limitations of our simulation work include the a need for understanding relevant associated fixed effect models; we focused on random effect models as they dominate modern repeatability studies. Lastly, a final limitation is the need to understand the role of changing the distance matrix for the non-parametric measures. Often, pseudo-distances, such as one minus Pearson correlation, are applied instead of Euclidean distances. However, this does not impact testing results if measurements are standardized and if measurements are non-negatively correlated. 

Another novel aspect of the manuscript is the establishment of the relationship between discriminability and rank sums. We emphasize that this relationship is between the testing statistics. The fact that the rank sum statistic utilizes more information than fingerprinting helps explain the power advantage of discriminability, which can be estimated using a linear transformation of rank sums. Therefore, we are confident in recommending discriminability over
fingerprinting in practice - unless there are concerns about mean shift batch effects. 
Similarly, rank sums share the performance properties of discriminability in most cases - once again except for mean shift batch effects. Nonetheless, we recommend using the full discriminability estimator over rank sums simply because it is more general.

\begin{figure}
    \centering
    \includegraphics[width=\linewidth]{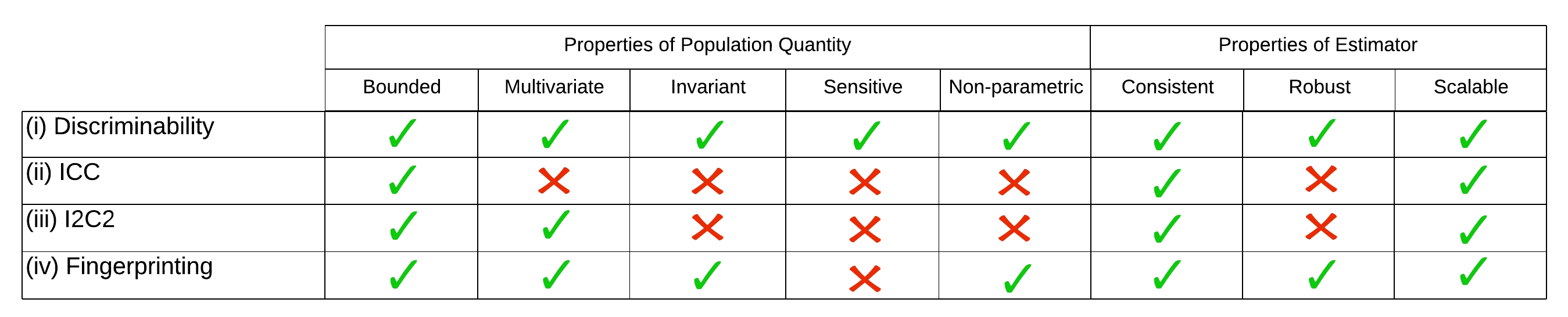}
    \caption{Comparison of data repeatability measures. 
    } 
    \label{fig:usability}
\end{figure}

Lastly, we compare the data repeatability measures in eight potentially useful properties (Figure \ref{fig:usability}).  We identify the eight usability properties as follows: 
\begin{enumerate}
    \item Bounded: The population quantity is defined on a fixed interval.
    \item  Multivariate: The population quantity is defined for data that is multivariate.
    \item Invariant: The statistic is agnostic to linear transformations of the measurements; that is, it is a unit-free quantity.
    \item Sensitive: The statistic is sensitive to changes in the reproducibility of the measurements.
    \item Nonparametric: The statistic does not make any parametric assumptions about the distribution of the data.
    \item Consistent: An estimator exists which is consistent for its associated population quantity.
    \item Robust: The statistic is robust to outliers with respect to the assumed model. 
    \item Scalable: The estimator of the population quantity can be computed in at most polynomial time.
\end{enumerate}
The boundedness and consistency properties might be violated if one directly summed up some of the between-measurement distances, instead of the nonparametric ranks or indicators, as the statistic. Of course, the non-parametric measurements are more robust relative to ICC and I2C2,
which tacitly rely on distributional assumptions.  This table outlines our thinking for
recommending discriminability as a default approach for measuring repeatability.

\paragraph{Acknowledgments}

This work was partially supported by funding from Microsoft Research.

 \bibliographystyle{apalike}

\clearpage

\bibliography{database}

\clearpage


\appendix

\section{Unbiasedness and Consistency of Discriminability Estimation}  \label{appendix:discrim}

Assume that for each individual $i$, we have $s$ repeated measurements. We define the local discriminability:
\begin{align}
    \hat D_{i, t, t'}^n &= \frac{\sum\limits_{i' \neq i} \sum\limits_{t'' = 1}^s \indicator{\delta_{i, t, t'} < \delta_{i, i', t, t''}}}{s\cdot (n - 1)}
\end{align}

where $\indicator{\cdot}$ is the indicator function, and $n$ is the total number of subjects. Then $\hat D_{i, t, t'}$ represents the fraction of observations from other subjects that are more distant from $\pmb x_{it}$ than $\pmb x_{it'}$, or a local estimate of the discriminability for individual $i$ between measurements $t$ and $t'$. The sample discriminability estimator is:
\begin{align}
    \hat D_{n} &= \frac{\sum\limits_{i = 1}^n \sum\limits_{t = 1}^s \sum\limits_{t' \neq t} \hat D_{i, t, t'}}{n\cdot s\cdot (s - 1)}
    \label{eqn:discr_est}
\end{align}

Where $D_{i, t, t'}$ is the local discriminability. We establish first the unbiasedness for the local discriminability, under the additive noise setting: 
\begin{align}
    \pmb x_{it} = \pmb v_i + \pmb \epsilon_{it} 
\end{align}
where $\pmb \epsilon_{it} \distas{ind} f_\epsilon$, and $\var{\pmb \epsilon_{it}} < \infty$ with $\expect{\epsilon_{it}} = c$. That is, our additive noise can be characterized by bounded variance and fixed expectation, and our noise is independent across subjects.

\begin{lemman}[local discriminability is unbiased for discriminability]
~
For fixed $n$:
\begin{align}
    \expect{\hat D_{i,t,t'}^n} = D
\end{align}
 that is; the local discriminability is unbiased for the true discriminability.
\label{lem:rdf_unb}
\end{lemman}

\begin{proof}
\begin{align*}
    \expect{\hat D_{i,t,t'}^n} &= \expect{\frac{\sum\limits_{i' \neq i} \sum\limits_{t'' = 1}^s \indicator{\delta_{i, t, t'} < \delta_{i, i', t, t''}}}{s\cdot (n - 1)}} \\
    &= \frac{\sum\limits_{i' \neq i} \sum\limits_{t'' = 1}^s\expect{\indicator{\delta_{i,t,t'} < \delta_{i,i',t,t''}}}}{s\cdot (n - 1)} \undereq \textrm{Linearity of Expectation}\\
    &= \frac{\sum\limits_{i' \neq i} \sum\limits_{t'' = 1}^s\prob{\delta_{i,t,t'} < \delta_{i,i',t,t''}}}{s\cdot (n - 1)} \\
    &= \frac{\sum\limits_{i' \neq i} \sum\limits_{t'' = 1}^s D}{s\cdot (n - 1)} \\
    &= \frac{\hcancel{s\cdot (n - 1)}\cdot D}{\hcancel{s\cdot (n - 1)}} \\
    &= D
\end{align*}
\end{proof}

Without knowledge of the distribution of $\pmb x_{it}$, we can instead estimate the discriminability via $\hat D(\pmb\phi, \pmb\psi)$, the observed sample discriminability. Consider the additive noise case. 
Recall that $\hat D_n \equiv \hat D_{n}(\pmb\phi, \pmb\psi)$, the sample discriminability for a fixed number of individuals $n$. We consider the following two lemmas:

\begin{lemman}[Unbiasedness of Sample Discriminability]
~
For fixed $n$:
\begin{align*}
    \expect{\hat{D}_n} = D
\end{align*}
that is; the sample discriminability is an unbiased estimate of discriminability.
\label{lem:discr_unb}
\end{lemman}

\begin{proof}
The proof of this lemma is a rather trivial application of the result in Lemma (\ref{lem:rdf_unb}).

Recall that sample discriminability is as-defined in Equation (\ref{eqn:discr_est}). Then:
\begin{align*}
    \expect{\hat D_n} &= \expect{\frac{\sum\limits_{i = 1}^n \sum\limits_{t = 1}^s \sum\limits_{t' \neq t} \hat D_{i, t, t'}}{n\cdot s\cdot (s - 1)}} \\
    &= \frac{\sum\limits_{i = 1}^n \sum\limits_{t =1}^s\sum\limits_{t' \neq t} \expect{\hat D_{i,t,t'}^n}}{n\cdot s\cdot (s - 1)} \\
    &= \frac{\sum\limits_{i = 1}^n \sum\limits_{t = 1}^s\sum\limits_{t' \neq t} D}{n\cdot s \cdot (s - 1)} \undereq \textrm{Lemma (\ref{lem:rdf_unb})} \\
    &= \frac{\hcancel{n\cdot s\cdot (s - 1)} \cdot D}{\hcancel{n\cdot s\cdot (s - 1)}}\\
    &= D
\end{align*}
\end{proof}

\begin{lemman}[Consistency of Sample Discriminability]
~
As $n \rightarrow \infty$:
\begin{align*}
    \hat D_n \convp[n \rightarrow \infty] D
\end{align*}
that is; the sample discriminability is a consistent estimate of discriminability.
\label{lem:discr_cons}
\end{lemman}

\begin{proof}
Recall that Chebyshev's inequality gives that:
\begin{align*}
    \prob{\abs*{\hat D_n - \expect{\hat D_n}} \geq \epsilon} &= \prob{\abs*{\hat D_n - D} \geq \epsilon} \undereq \textrm{$\hat D_{i,t,t'}^n$ is unbiased} \\
    &\leq \frac{\var{\hat D_n}}{\epsilon^2}
\end{align*}
To show convergence in probability, it suffices to show that $\var{\hat D_n} \xrightarrow[n \rightarrow \infty]{} 0$. Then:
\begin{align*}
    \var{\hat D_n} &= \var{\frac{\sum\limits_{i = 1}^n \sum\limits_{t = 1}^s\sum\limits_{t' \neq t} \hat D_{i,t,t'}^n}{n \cdot s \cdot (s - 1)}} \\
    &= \frac{1}{m_*^2}\var{\sum\limits_{i = 1}^n\sum_{t = 1}^s\sum_{t' \neq t}\sum\limits_{i' \neq i} \sum\limits_{t'' = 1}^s \indicator{\delta_{i, t, t'} < \delta_{i, i', t, t''}}},\;\;m_* = n\cdot s \cdot (s - 1)\cdot (n - 1)\cdot s \\
    &= \frac{1}{m_*^2}\sum_{i,i',t,t',t''}\sum_{j,j',r,r',r''}\cov{\indicator{\delta_{i, t, t'} < \delta_{i, i', t, t''}}, \indicator{\delta_{j, r, r'} < \delta_{j, j', r, r''}}}
\end{align*}

Note that there are, in total, $m^2_*$ covariance terms in the sums. For each term, by Cauchy-Schwarz:
\begin{align*}
    \abs*{\cov{\indicator{\delta_{i,t,t'} < \delta_{i,i',t,t''}}, \indicator{\delta_{i,t,t'} < \delta_{i,j',t,r''}}}} &\leq \sqrt{\var{\indicator{\delta_{i,t,t'} < \delta_{i,i',t,t''}}} \cdot \var{\indicator{\delta_{i,t,t'} < \delta_{i,j',t,r''}}}} \\
    &\leq \sqrt{\frac{1}{4}\cdot \frac{1}{4}} = \frac{1}{4}
\end{align*}
Furthermore, note that $\indicator{\delta_{i,t,t'} < \delta_{i,i',t,t''}} = f(\pmb x_{i,t}, \pmb x_{i,t'}, \pmb x_{i',t''})$. Under the assumption of between-subject independence, then $\indicator{\delta_{i,t,t'} < \delta_{i,i',t,t''}} \indep g\parens*{\pmb x_{i'', q}: i'' \neq i, i'}$, as it will be independent of any function $g(\cdot)$ of subjects other than $i$ and $i'$. Then as long as $\set*{i,i'} \cap \set*{j,j'} = \varnothing$, $\indicator{\delta_{i,t,t'} < \delta_{i,i',t,t''}} \indep \indicator{\delta_{j,r,r'} < \delta_{j,j',r,r''}}$. Under the assumption that $n_i = s \forall i$, $m_* = ns^2(s-1)(n-1)$. Then there are $(n-2)s^2(s-1)(n-3)$ combinations of $j,j',r,r',r''$ that will produce covariances taking values of $0$, and $m_* - (n-2)s(s-1)(n-3)s = (4n - 6)\cdot s^2\cdot(s-1)$ combinations that may be non-zero. Then:
\begin{align*}
    \var{\hat D_n} &=  \frac{1}{m_*^2}\sum\limits_{i,i',t,t',t''}\sum_{j,j',r,r',r''}\cov{\indicator{\delta_{i, t, t'} < \delta_{i, i', t, t''}}, \indicator{\delta_{j, r, r'} < \delta_{j, j', r, r''}}} \\
    &\leq \frac{\sum_{i,i',t,t',t''} (4n - 6)s^2(s-1)}{4m_*^2} \\
    &= \frac{(4n - 6)s^2(s-1)}{4ns^2(s-1)(n-1)} \\
    &= \frac{4n - 6}{4n(n-1)} \\
    &< \frac{1}{n} \\
    &\xrightarrow[n \rightarrow \infty]{} 0
\end{align*}
\end{proof}


\section{Consistency of the Alternative Discriminability Estimation} \label{appendix:alter}

Another sensible estimator of discriminability in the rank form is 
\begin{equation}
\Tilde{D} = \frac{n^2s^2(s-1) - \sum_{t=1}^s \sum_{t'\not = t} \sum_{i=1}^n r_{ii}^{tt'} }{ns(s-1)(n-1)s}
\end{equation}
or $\Tilde{D} - \frac{s-2}{2(n-1)s}$, where
\begin{align}
\label{eq:compare}
\Tilde{D} - \hat D 
& \leq 
\frac{ns(s-1)(s-2)}{2ns(s-1)(n-1)s} \\ 
& = 
\frac{s-2}{2(n-1)s}. 
\end{align}
Equality is taken in (\ref{eq:compare}) when no tie exists between $\delta_{i,i}^{t,t'}$ and $\delta_{i,i}^{t,t''}$ for all $i \in \{1, \dots, n\}$, $t 
\in \{1, \dots s \}$, $t' \not\in \{t\}$, $t'' \not\in \{t, t'\}$. Therefore we have that $\Tilde{D}$ and $\Tilde{D} - \frac{s-2}{2(n-1)s}$ are also consistent estimators for discriminability. In fact $\hat{D} = \Tilde{D} - \frac{s-2}{2(n-1)s}$ when assuming complex continuous measurements with no ties in distance ranking. 

\section{Convergence of Fingerprint Index} \label{appendix:discrim_fp}

Note that the definition of fingerprint index, (\ref{eq:FI}), can be rewritten: 
\begin{align*}
    F_{index} = & \prob{\delta_{i, 1, 2} < \delta_{i, i', 1, 2}; \forall i' \not = i}\\ 
    = & \prob{\sum_{i' \not = i}\indicator{\delta_{i, 1, 2} < \delta_{i, i', 1, 2}} = n-1} \\
    \stackrel{def}{=} & \prob{W = n-1}, 
\end{align*}
where $W \stackrel{def}{=} \sum_{i' \not = i}\indicator{\delta_{i, 1, 2} < \delta_{i, i', 1, 2}}$ is a sum of $n-1$ possibly correlated Bernoulli variables. 
Define the conditional discriminability given $\pmb x_{i} = (\pmb{x}_{it}: t = 1, 2)$ as 
\begin{align*}
    D(\pmb x_i) = \prob{\delta_{i, 1, 2} < \delta_{i, i', 1, 2} | \pmb x_i}, 
\end{align*}
where $i' \not = i$. Then by iterated expectation we have 
\begin{align*}
    F_{index}
    = & \ex \ex(\indicator{W = n-1} | \pmb x_i) \\
    = & \ex(D(\pmb x_i)^{n-1}), 
\end{align*}
because given any realization of $\pmb x_i$, $\indicator{\delta_{i, 1, 2} < \delta_{i, i', 1, 2}}, i' = 1, \dots, i-1, i+1, \dots, n$, are conditional IID , and thus $W | \pmb x_i$ follows a Binomial distribution. Note that $|D(\pmb x_i)| < 1$. By the Dominated Convergence Theorem for random variables, under a regularity condition that $D(\pmb x_i)^{n-1} \overset{a.s.}{\to} 0$ we have 
\begin{align*}
    F_{index} = \ex(D(\pmb x_i)^{n-1}) \to 0. 
\end{align*}
With the existence of random measurement errors, in general we have that $\prob{D(\pmb x_i) = 1} \overset{a.s.}{=} 0$, which implies $\prob{\lim_{n \to \infty} D(\pmb x_i)^{n-1} = 0} = 1$. 

However, the relationship between the $n-1$-th order moment, $F_{index} = \ex(D(\pmb x_i)^{n-1})$, and the first order moment, $D = \ex(D(\pmb x_i))$, is not solely decided by the correlations between $\prob{\indicator{\delta_{i, 1, 2} < \delta_{i, i', 1, 2}} | \pmb x_i}$ across distinct $i'$, and therefore is dependent on the actual distribution of data $\pmb x_i$.

\section{Discriminability and I2C2} \label{appendix:mono}

We will give the approximation and then prove the non-decreasing bounds in Section \ref{sec:anova_approx}. 

Applying the Satterthwaite approximation that matches the first two moments
\citep{yuan2010two}, we have 
$\sum_{u = 1}^r \lambda_u U_u
\stackrel{D}{\approx} g_1 \chi^2_{h_1}$ and 
$\sum_{u = r + 1}^{r + w} \lambda_u' U_u
\stackrel{D}{\approx} g_2 \chi^2_{h_2}$, 
where 
\begin{align*}
 g_1 &= \left (\sum_{u = 1}^r \lambda_u^2 \right ) / \left (\sum_{u = 1}^r \lambda_u \right ), \\   
 h_1 &= \left (\sum_{u = 1}^r \lambda_u \right )^2 / \left (\sum_{u = 1}^r \lambda_u^2 \right ), \\
 g_2 &= \left (\sum_{u = r + 1}^{r + w} \lambda_u'^2 \right ) / \left (\sum_{u = r + 1}^{r + w} \lambda_u' \right ), \\
 h_2 &= \left (\sum_{u = r + 1}^{r + w} \lambda_u' \right )^2 / \left (\sum_{u = r + 1}^{r + w} \lambda_u'^2 \right ). 
\end{align*}
Let $V_1 = \sum_{u = 1}^r \lambda_u = h_1g_1$, $W_1 = \sum_{u = 1}^r \lambda_u^2$, $V_2 = \sum_{u = r + 1}^{r+w} \lambda_u' = h_2g_2$, $W_2 = \sum_{u = r+1}^{r+w} \lambda_u'^2$. Thus: 
\begin{align}
D = P(Z \leq 0)
& \approx P\left(\frac{g_1\chi^2_{h_1}}{g_2\chi^2_{h_2}} \leq 1\right) \nonumber\\
& = P\left(\frac{\chi^2_{h_1}/h_1}{\chi^2_{h_2}/h_2} \leq \frac{h_2g_2}{h_1g_1}\right) \nonumber\\ 
& = F_{F\left(\frac{V_1^2}{W_1}, \frac{V_2^2}{W_2}\right)}\left(\frac{V_2}{V_1}\right). \nonumber
\end{align}
Here, $\frac{\chi^2_{h_1}/h_1}{\chi^2_{h_2}/h_2}$ follows $F$ distribution with degrees of freedom $h_1 = \frac{V_1^2}{W_1}, h_2 = \frac{V_2^2}{W_2}$. 


Now we derive the non-decreasing bounds. 

Note that
\begin{align*}
\pmb H
\stackrel{def}{=}
\left(
\begin{array}{cc}
2\pmb \Sigma & -\pmb \Sigma \\ 
\pmb \Sigma & -2\pmb \Sigma - 2\pmb \Sigma_{\pmb \mu}
\end{array}
\right)
=
\left(
\begin{array}{cc}
2\pmb \Sigma & \pmb \Sigma \\ 
\pmb \Sigma & 2\pmb \Sigma + 2\pmb \Sigma_{\pmb \mu}
\end{array}
\right)
\left(
\begin{array}{cc}
\pmb I & \pmb 0 \\ 
\pmb 0 & -\pmb I
\end{array}
\right)
\stackrel{def}{=}
\pmb P\pmb M
\end{align*}
is congruent to $\pmb M = \left( \begin{array}{cc}
\pmb I & \pmb 0 \\ 
\pmb 0 & -\pmb I
\end{array}
\right)
$ since $\pmb P$ is symmetric and positive definite. 
By Sylvester's law of inertia \citep{sylvester1852xix} we have $r = w = l$, i.e. the numbers of positive and negative eigenvalues of $\pmb H$ are both $l$. 

Denote the sums of positive or negative eigenvalues of the matrix $\pmb H$ as $\sigma_+(\pmb H)$ or $\sigma_-(\pmb H)$, respectively. We will apply the monotonicity of $\sigma_\pm(\pmb H) = \sigma_\pm(\pmb M \pmb P)$ \citep{lieb1991convexity} for the following statements (Appendix \ref{appendix:mono}): 
\begin{lemman}[Monotonicity of Sums of Positive or Negative Eigenvalues]
\label{lem:mono}
\begin{align}
    tr(\frac{3}{2}\pmb\Sigma + 2\pmb\Sigma_{\pmb\mu}) \leq |\sigma_-(\pmb H)| & = V_2 \leq tr(2\pmb\Sigma + 2\pmb\Sigma_{\pmb\mu}) \label{eq:example} \\ 
    tr(\frac{3}{2}\pmb\Sigma) \leq \sigma_+(\pmb H) & = V_1 \leq tr(2\pmb\Sigma) \label{eq:example1}
. 
\end{align}
\end{lemman}

\begin{proof}
For (\ref{eq:example}), note that $\pmb P - 
\left(
\begin{array}{cc}
\pmb 0 & \pmb 0 \\ 
\pmb 0 & 2\pmb\Sigma_{\pmb\mu} + v\pmb\Sigma
\end{array}
\right)
= 
\left(
\begin{array}{cc}
2 & 1 \\ 
1 & 2 - v
\end{array}
\right)
\otimes \pmb \Sigma
$ is positive definite for all $v \in (0, 3/2)$. Therefore 
\begin{align*}
|\sigma_-(\pmb M \pmb P)| \geq |\sigma_-(\pmb M
\left(
\begin{array}{cc}
\pmb 0 & \pmb 0 \\ 
\pmb 0 & 2\pmb\Sigma_{\pmb\mu} + v\pmb\Sigma
\end{array}
\right)
)|
\end{align*}
for all $v \in (0, 3/2)$. 
Finally, 
\begin{align*}
    |\sigma_-(\pmb M \pmb P)| \geq \lim_{v \to \frac{3}{2}} 
    |\sigma_-(\pmb M
    \left(
\begin{array}{cc}
\pmb 0 & \pmb 0 \\ 
\pmb 0 & 2\pmb\Sigma_{\pmb\mu} + v\pmb\Sigma
\end{array}
\right)
)|
=
\trace{\frac{3}{2}\pmb\Sigma + 2\pmb\Sigma_{\pmb\mu}}. 
\end{align*}
Meanwhile 
$ 
\left(
\begin{array}{cc}
v\pmb \Sigma & \pmb \Sigma \\ 
\pmb 0 & 2\pmb\Sigma_{\pmb\mu} + v\pmb\Sigma
\end{array}
\right) - \pmb P
= 
\left(
\begin{array}{cc}
v - 2 & 0 \\ 
-1 & v - 2
\end{array}
\right)
\otimes \pmb \Sigma
$ is positive definite for all $v > 2$. Similarly, 
\begin{align*}
    tr(2\pmb\Sigma_{\pmb\mu} + 2\pmb\Sigma) 
    = \lim_{v \to 2} 
    |\sigma_-(
    \pmb M\left(
\begin{array}{cc}
v\pmb \Sigma & \pmb \Sigma \\ 
\pmb 0 & 2\pmb\Sigma_{\pmb\mu} + v\pmb\Sigma
\end{array}
\right)
    )| 
    \geq 
    |\sigma_-(\pmb M \pmb P)|. 
\end{align*}

To get (\ref{eq:example1}) from (\ref{eq:example}), note $V_1 - V_2 = tr(\pmb H) = tr(-2\pmb\Sigma_{\pmb \mu})$. 
\end{proof}

Therefore, 
\begin{align*}
    \frac{V_2}{V_1} = 1 + \frac{2tr(\pmb \Sigma_{\pmb\mu})}{V_1} 
    \in \left(1 + \frac{tr(\pmb \Sigma_{\pmb\mu})}{tr(\pmb \Sigma)} , 
1 + \frac{4}{3} \cdot \frac{tr(\pmb \Sigma_{\pmb\mu})}{tr(\pmb \Sigma) } \right)
= 
\left( f_1(\frac{tr(\pmb \Sigma_{\pmb\mu})}{tr(\pmb \Sigma) + tr(\pmb \Sigma_{\pmb\mu})}), 
f_2(\frac{tr(\pmb \Sigma_{\pmb\mu})}{tr(\pmb \Sigma) + tr(\pmb \Sigma_{\pmb\mu})})
\right ), 
\end{align*}
where $f_1(v) = 1 + v/(1-v)$ and $f_2(v) = 1 + (4/3) \cdot v/(1-v)$ are both non-decreasing functions. 


If $l = 2$, by the monotonicity of F distribution \citep{ghosh1973some} we have bounds for the approximation (\ref{eq:approxF}):
\begin{align*}
F_{F(2, 1)}\left(f_1(\frac{tr(\pmb \Sigma_{\pmb\mu})}{tr(\pmb \Sigma) + tr(\pmb \Sigma_{\pmb\mu})})\right) 
    \leq F_{F(2, 1)}\left(\frac{V_2}{V_1}\right) \leq 
    F_{F\left(\frac{V_1^2}{W_1}, \frac{V_2^2}{W_2}\right)}\left(\frac{V_2}{V_1}\right) 
    \leq F_{F(1, 2)}\left(f_2(\frac{tr(\pmb \Sigma_{\pmb\mu})}{tr(\pmb \Sigma) + tr(\pmb \Sigma_{\pmb\mu})})\right)
    , 
\end{align*}
where $f_1$, $f_2$, $F_{F(2, 1)}$, $F_{F(1, 2)}$ are all non-decreasing functions. 

For $l \geq 3$, when the dispersion measures $V_1^2/W_1$ and $V_2^2/W_2$ remain constants (in fact $1 \leq V_j^2/W_j \leq l$ for $j = 1, 2$ by the property of $l_1$ and $l_2$ norms), the approximation of $D$ in (\ref{eq:approxF}) is bounded by a non-decreasing interval of I2C2 (Figure \ref{fig:bound_dispersion}): 
\begin{align*}
 F_{F\left(\frac{V_1^2}{W_1}, \frac{V_2^2}{W_2}\right)}\left(
 f_1(\frac{tr(\pmb \Sigma_{\pmb\mu})}{tr(\pmb \Sigma) + tr(\pmb \Sigma_{\pmb\mu})})
 \right)
\leq 
    F_{F\left(\frac{V_1^2}{W_1}, \frac{V_2^2}{W_2}\right)}\left(\frac{V_2}{V_1}\right)
    \leq 
    F_{F\left(\frac{V_1^2}{W_1}, \frac{V_2^2}{W_2}\right)}\left(
    f_2(\frac{tr(\pmb \Sigma_{\pmb\mu})}{tr(\pmb \Sigma) + tr(\pmb \Sigma_{\pmb\mu})})
    \right).
\end{align*}

\end{document}